\documentclass[letterpaper]{article} 
\usepackage{aaai25}  
\usepackage{times}  
\usepackage{helvet}  
\usepackage{courier}  
\usepackage[hyphens]{url}  
\usepackage{graphicx} 
\usepackage{natbib}  
\usepackage{caption} 
\usepackage{newfloat}
\usepackage{listings}
\DeclareCaptionStyle{ruled}{labelfont=normalfont,labelsep=colon,strut=off} 
\usepackage[linesnumbered,ruled,vlined]{algorithm2e} 

\usepackage{algpseudocode}
\SetKwInput{KwInput}{Input}                
\SetKwInput{KwOutput}{Output}  
\usepackage{array}
\usepackage{multirow}
\usepackage{booktabs}
\usepackage{comment}
\usepackage{amsfonts}
\usepackage{mathtools}

\newcommand{\spara}[1]{\smallskip\noindent{\bf #1}}
\newcommand{\ourSolution}{\texttt{IOHunter}}
\graphicspath{{./figs/}}

\title{\ourSolution: Graph Foundation Model to Uncover Online Information Operations}
\author{
    Marco Minici\textsuperscript{\rm 1,\rm 2},
    Luca Luceri\textsuperscript{\rm 3,\rm 4},
    Francesco Fabbri\textsuperscript{\rm 5},
    Emilio Ferrara\textsuperscript{\rm 3,\rm 4}
}
\affiliations{
    \textsuperscript{\rm 1}ICAR-CNR, Rende, Italy\\
    \textsuperscript{\rm 2}University of Pisa, Pisa, Italy\\
    \textsuperscript{\rm 3}Information Sciences Institute, University of Southern California, Marina Del Rey, CA, USA\\
    \textsuperscript{\rm 4}Thomas Lord Department of Computer Science, University of Southern California, Los Angeles, CA, USA\\
    \textsuperscript{\rm 5}Spotify, Barcelona, Spain\\
    marco.minici@icar.cnr.it,
    lluceri@isi.edu,
    francescof@spotify.com,
    emiliofe@usc.edu
}

\begin{document}

\maketitle

\begin{abstract}
Social media platforms have become vital spaces for public discourse, serving as modern agorás where a wide range of voices influence societal narratives. However, their open nature also makes them vulnerable to exploitation by malicious actors, including state-sponsored entities, who can conduct information operations (IOs) to manipulate public opinion.
The spread of misinformation, false news, and misleading claims threatens democratic processes and societal cohesion, making it crucial to develop methods for the timely detection of inauthentic activity to protect the integrity of online discourse. In this work, we introduce a methodology designed to identify users orchestrating information operations, a.k.a. IO drivers, across various influence campaigns. Our framework, named IOHunter, leverages the combined strengths of Language Models and Graph Neural Networks to improve generalization in supervised, scarcely-supervised, and cross-IO contexts. Our approach achieves state-of-the-art performance across multiple sets of IOs originating from six countries, significantly surpassing existing approaches. This research marks a step toward developing Graph Foundation Models specifically tailored for the task of IO detection on social media platforms.
\end{abstract}

\begin{links}
    \link{Code}{https://github.com/mminici/SocGFM}     \link{Datasets}{https://zenodo.org/records/13357621}
\end{links}

\section{Introduction}
Online social media platforms have become essential for fostering public discourse, where users engage in debates on critical political and social issues. The integrity of these online spaces is paramount, given their significant role in shaping public opinion and influencing societal outcomes, such as elections or public health interventions \cite{starbird2019disinformation,ferrara2015manipulation, nogara2022disinformation}. However, these platforms are increasingly vulnerable to state-sponsored Information Operations (IOs), which seek to manipulate narratives, spread disinformation, and foster division through the promotion of hate speech and other harmful content \cite{badawy2018analyzing,zannettou2019disinformation,suresh2023tracking, Minici_Cinus_Luceri_Ferrara_2024}. The proliferation of such campaigns poses a significant threat to democratic processes, highlighting the urgent need for robust methods to detect and mitigate these operations \cite{wef2024globalrisks}.

The development of machine learning techniques for IO detection is a rapidly growing area of research. Recent studies, such as \citeauthor{luceri2024unmasking}~\citeyear{luceri2024unmasking}, have demonstrated the potential to leverage graph machine learning techniques for this purpose. Specifically, they leverage node2vec embeddings of similarity networks 
constructed from behavioral traces, such as co-sharing patterns, to detect coordinated users driving IOs, namely \textit{IO drivers}.
Their findings highlight the potential of using topological structures based on similarity patterns, combined with graph machine learning techniques, for detecting IOs. 
However, they did not explore whether recent advancements in Graph Neural Networks (GNNs) could further improve performance. 
GNN-based approaches not only provide a more powerful framework for modeling online user behavior but also offer inductive capabilities, enabling generalization to nodes not seen during training. This flexibility is particularly crucial for deploying auditing tools in dynamic environments, where threats can arise from new users or, even more critically, from IOs originating in different geopolitical contexts.

Generalizing across different IOs—referred to here as \textit{cross-IO detection}—is inherently difficult, 
as distinct IOs often employ different coordination strategies and may operate in different languages~\cite{luceri2024unmasking}. 

\spara{Contribution of this work.}
In this paper, we propose \ourSolution{}, an architecture for IO detection that combines the message-passing paradigm of GNNs with multi-modal information derived from both network structure and textual content.
Unlike existing approaches that are either based on graph structure or textual content, \ourSolution{} builds on the emerging concept of Graph Foundation Models (GFMs). Traditional GNNs are typically trained from scratch on specific tasks and datasets, limiting their ability to generalize across different domains. In contrast, \ourSolution{} integrates GNNs with embeddings extracted from Language Models to create a GFM capable of leveraging large-scale, diverse graph data with the goal of rapidly adapting to new tasks or datasets.
Our approach is thoroughly detailed in the Methodology section. 
We evaluate \ourSolution{} on six datasets from Twitter, each representing an IO originating from distinct geopolitical contexts: UAE, Cuba, Russia, Venezuela, Iran, and China. \ourSolution{} achieves improvements of up to +20\% in Macro-F1 compared to the state-of-the-art in IO detection.
Additionally, we demonstrate the robustness of \ourSolution{} in scenarios with limited data availability and its effectiveness in cross-IO detection tasks when pretrained and fine-tuned with minimal labeled data. 

\section{Related Work}

\subsection{Machine Learning-Based IO Detection}

Research in IO detection has extensively analyzed individual account activities to detect participation in influence campaigns, particularly focusing on bots (software-controlled accounts) and trolls (state-backed human operators) \cite{mazza2022investigating, ferrara2023social}. Bot detection has been a focal point, with various solutions utilizing machine learning strategies to \textit{(i)} identify bot characteristics, such as posting frequency, content patterns, and network behavior \cite{Yang_2019, chen2018unsupervised, cresci2016dna}, and/or \textit{(ii)} distinguish patterns of bot behavior from organic human behavior \cite{pozzana2020measuring}. Notably, the Botometer tool \cite{Yang_2019, yang2022botometer} has played a significant role in scaling bot activity research on Twitter, enabling studies focused on the identification of bot-driven influence campaigns \cite{shao2018spread, stella2018bots, deb2019perils, grinberg2019fake, luceri2019evolution}.

However, recent studies have emphasized that IO coordination extends beyond automated bots, highlighting the role of human-operated trolls in these operations \cite{nizzoli2021coordinated, Hristakieva_2022}. 
Research on state-sponsored trolls has been categorized into three primary detection methods: content-based, behavioral-based, and sequence-based approaches. Content-based methods analyze the linguistic features of posts to identify deceptive or coordinated messaging \cite{alizadeh2020content, luceri2024leveraging, im2020still}. Behavioral-based approaches focus on user activity patterns, such as posting activity and interaction signals, to detect coordinated inauthentic behavior \cite{luceri2020detecting, kong2023interval, sharma2021identifying}. Sequence-based techniques, on the other hand, model the temporal sequence of actions to uncover orchestrated activities over time \cite{nwala2023language, ezzeddine2022characterizing}.

\subsection{Network-Based IO Detection}

In addition to machine learning-based methods, a significant body of research has focused on detecting IOs through network-based approaches. These methods aim to uncover tactics of online coordination by identifying unexpected or exceptional similarities in the actions of multiple users \cite{Pacheco_2020, Pacheco_2021, nizzoli2021coordinated, mannocci2024detection, magelinski2022synchronized, luceri2024unmasking}. The underlying assumption is that connections between highly similar users—such as those who share the same content, use similar hashtags, or post at synchronized times—can reveal coordinated clusters likely engaged in IOs.

Network-based detection typically involves constructing networks that represent user similarities using edge weights, where higher weights indicate stronger behavioral correlations. By exploiting network properties to filter out organic users, researchers identify clusters of users exhibiting collective similarity and potentially driving IOs \cite{Pacheco_2021, luceri2024unmasking}. This approach has proven effective in revealing coordinated activities, providing valuable insights into the structure and scale of influence campaigns.

\subsection{Graph Foundation Model}
The challenge of training models capable of generalizing across diverse graph domains and tasks has recently attracted significant attention~\cite{mao2024graph}. Earlier work in this area has focused on self-supervised approaches to enable rapid adaptation to downstream tasks on the same graph~\cite{lu2021learning, de2024personalized} or to generalize across graphs from different domains~\cite{qiu2020gcc, jiang2021pre}. However, these methods do not address the challenge of integrating multi-modal information.  

Recent advancements have explored the use of LMs to generate transferable features in heterogeneous graph settings, such as personalization~\cite{damianou2024towards} and e-commerce applications~\cite{xie2023graph}. Others, like PRODIGY~\cite{huang2024prodigy} and OFA~\cite{liuone}, focus on adapting graph tasks to leverage the generalization capabilities of LLMs for in-context learning tasks.  

In contrast, our approach specifically targets the problem of IO detection across three distinct learning regimes. Within this context, we demonstrate that integrating multi-modal signals combined with massive pre-training—while keeping the LM weights frozen—is an effective strategy to achieve a GFM tailored to the IO detection task.



\section{Problem Definition}
We are given an undirected graph $G = (V, E)$, where $V = \{v_1, ..., v_n\}$ represents the set of nodes, and $E \subset V \times V$ denotes the set of edges. In this context, $G$ models the relationships between social media users, with an edge $\ell(v_1, v_2) \in E$ existing if two users $v_1$ and $v_2$ are considered similar. 
We will also refer to $G$ as the similarity network\footnote{We use network and graph interchangeably.}.
For each user $v_i \in V$, we have access to a set of content $C_{i}$ (e.g., texts, images) that $v_i$ has shared on the social network. Additionally, each user $v_i$ is associated with a label $y_i \in \{0, 1\}$, where $1$ indicates an \textit{IO driver} and $0$ represents a legitimate user.

Our objective is to learn two functions: a multi-modal projection $p_{\psi} : V \rightarrow \mathbb{R}^d$ and a probabilistic node classifier $f_{\theta} : \mathbb{R}^d \rightarrow [0,1]$. The multi-modal projection $p_{\psi}$ maps each node $v_i$ to a point $z_i = p_{\psi}(v_i \mid G, C_i)$ in a $d$-dimensional latent space, utilizing both the contextual information from $G$ and the content $C_i$ shared by the user. For simplicity, we will refer to this embedding as $p_{\psi}(v_i)$.

The node classifier $f_{\theta}$ takes the low-dimensional representation $z_i$ as input and outputs a score $s_i = f_{\theta}(z_i)$, indicating the likelihood that $v_i$ is an \textit{IO driver}. Given that our task is a binary classification problem, we optimize the complete set of learnable parameters $\Theta = \{\psi, \theta\}$ by minimizing the Binary Cross-Entropy loss for each node $v_i$ as follows:
\begin{equation}
    \mathcal{L}(\Theta, v_i) = -\left[y_i \log(s_i) + (1-y_i)\log(1-s_i)\right].
\end{equation}

\section{Methodology}
\begin{figure*}
    \centering
    \includegraphics[width=0.9\linewidth]{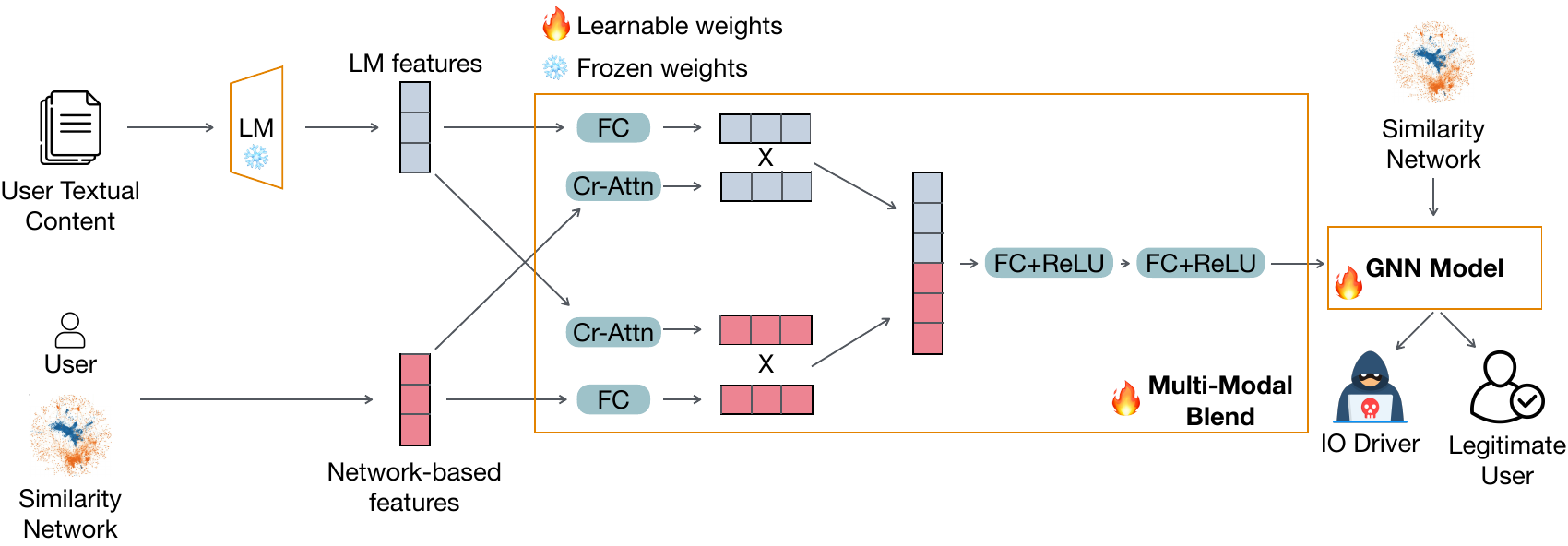}
    \caption{Illustration summarizing \ourSolution{}. We feed the user posts to a multi-lingual SBert and then average all post embeddings to obtain a unique textual representation. SBert is frozen and not optimized. We also extract a degree-based embedding from the fused similarity network. Both modality embeddings are blended through a cross-attention layer and a couple of fully connected layers after concatenation. The obtained multi-modal representation is then fed to a GNN model that determines whether the user is an IO Driver or a legitimate user. Both the multi-modal embedding module and the GNN model are optimized during the training phase.}
    \label{fig:framework}
\end{figure*}

Our objective is to detect \textit{IO drivers} by integrating two sources derived from the behavioral traces of social media users: \textit{(i)} the textual content they share and \textit{(ii)} the similitude of their sharing activities as captured by similarity networks.
Our approach, illustrated in Fig. \ref{fig:framework}, begins by extracting textual embeddings from user-shared posts and graph embeddings from the \textit{Fused Similarity Network} (see below). 
These two data modalities are then blended using a cross-attention module, allowing the model to learn interactions between the content and network contexts. 
The resulting multi-modal embeddings are subsequently input into a GNN, which leverages this enriched representation to accurately predict user categories.

\spara{Fused Similarity Network.}
The process of constructing a similarity graph generally follows a consistent approach in the literature \cite{Pacheco_2020, Pacheco_2021, luceri2024unmasking}.
We consider different similarities --- including the sharing of identical links (\textit{co-URL}), hashtags (\textit{co-hashtag}), or content (\textit{text similarity}), the re-sharing of the same tweets (\textit{co-retweet}), and automation-driven actions such as rapid retweeting (\textit{fast-retweet}) --- to build five distinct similarity networks.
We construct a bipartite graph between users and entities, where the entities correspond to the specific behavioral trace being analyzed (e.g., for the Co-URL trace, the entities are the URLs). 
The Fast Retweet bipartite network is constructed similarly to the co-retweet bipartite network, but it excludes connections where the retweet takes more than 10 seconds.
In this bipartite network, users are linked to entities based on their sharing activities, with weights assigned using TF-IDF to account for the popularity of each entity. Consequently, each user is represented as a TF-IDF vector of the shared entities. This bipartite graph is then transformed into a similarity network, where users are connected based on the similarity of their behavioral traces.

Building on \cite{luceri2024unmasking}, we combine the five similarity networks by linking two nodes in a \textit{Fused Similarity Network} if they are connected in
any of the individual similarity networks.

\spara{Content Embedding.}
For each user $v_i$, we extract a low-dimensional embedding from its set of shared content $C_i$. 
Since the main source of information in our experiments is textual content, we use a Sentence Transformer, SBert~\cite{reimers-2019-sentence-bert}.
We refer to $c_i \in \mathbb{R}^{d_c}$ as the content embedding of $v_i$, where $d_c$ is output dimensionality of SBert.
If $\lvert C_i \rvert > 1$ then $c_i$ will be the average of the embeddings of each single content in $C_i$.

\spara{Contextual Embedding.}
To consider the contextual information provided by the graph $G$, we abide by the best practice ~\cite{cui2022positional}, and divide degree values into $d_g$ buckets, then map the degree value distributed in each bucket range into one class, and finally
construct a unique one-hot vector for each class.
Hence, we encode the structural information of each node $v_i$ to a one-hot vector $g_i \in \{0,1\}^{d_g}$.

\spara{Multi-Modal Blend.}
Given the content and contextual embeddings $c_i$ and $g_i$, we now need to merge them in a unique projection $z_i$ that can be used by a node classifier to detect if the node $v_i$ is an \textit{IO driver}.
We propose to use a cross-attention mechanism, loosely inspired by~\cite{abavisani2020multimodal}. The main advantage of this approach is to filter out any irrelevant information that might come from the other modality, thus allowing to learn how to best combine content and contextual information.

First, we remap both $c_i$ and $g_i$ to a $d-$dimensional space using a fully-connected layer and a $ReLU$ non-linear activation as follows:
\begin{align}
    \tilde{c}_i = ReLU(W_c^Tc_i + b_c), \\
    \tilde{g}_i = ReLU(W_g^Tg_i + b_g).
\end{align}
We also compute the cross-attention coefficients $\alpha_c, \alpha_g$ for each modality:
\begin{align}
    \alpha_c = ReLU({W'_c}^Tg_i + b'_c), \\
    \alpha_g = ReLU({W'_g}^Tc_i + b'_g).
\end{align}
The node representation $z_i$ is obtained by an element-wise multiplication of cross-attention coefficients and content/contextual embeddings, followed by a concatenation operation:
\begin{equation}
    z_i = \alpha_c\odot \tilde{c}_i \mathbin\Vert \alpha_g\odot \tilde{g}_i.
\end{equation}

The final node representation $z_i$ is further refined by two fully-connected layers $W^{(1)}_z, b^{(1)}_z, W^{(2)}_z, b^{(2)}_z$, each one followed by a non-linear ReLU function.

We will refer to all these steps using the function $z_i = p_{\psi}(v_i)$, following the notations used in the \textit{Problem Definition} section.

\spara{GNN Model.}
The derived multi-modal node embedding $z_i = p_{\psi}(v_i)$ is given as input to a function $f_{\theta}$ that outputs the probability of $v_i$ to be an \textit{IO driver}.
We use a generic GNN model~\cite{wu2020comprehensive} (with a final logit head) to model $f_{\theta}$, given the state-of-the-art performances of these models on graph-related tasks. Specifically, we leverage GCN~\cite{kipf2022semi} and Sage~\cite{hamilton2017inductive} models.

\smallskip
The overall architecture is graphically described in Figure~\ref{fig:framework}, while the the end-to-end learning procedure is shown in Algorithm~\ref{alg:iobuster}.

\begin{algorithm}[t]
\DontPrintSemicolon

  \KwInput{Graph $G=(V, E)$, \\
  $\quad\quad\quad$User Content $C = \{C_i \mid \forall v_i \in V\}$ \\
  }
  \KwResult{Model parameters $\Theta$}
  Compute $c_i$ and $g_i$ for all users $v_i \in V$ \\
  Initialize $\Theta^{(0)}$\\
  \While{training has not converged}
   {
   Sample $V_{\text{batch}} \subseteq V$\\
   $l_{\text{tot}} = 0$ \\
   \ForAll{$v_i \in V_{\text{batch}}$}{
        $l_{\text{tot}} \mathrel{\scriptstyle+=} \mathcal{L}(\Theta^{(t)}, v_i)$ \tcp*{eq. 1}
   }
   $\Theta^{(t+1)} \leftarrow \texttt{BackPropagate}(l_{\text{tot}})$ \tcp*{Update model}
    }
\caption{Learning Procedure of \ourSolution{}}
\label{alg:iobuster}
\end{algorithm}

\section{Experiments}
This experimental section examines whether our \ourSolution{} can detect IOs on social media networks. 
We address the following research questions investigating the applicability of our proposal in different, realistic data regimes:

\begin{itemize}
    \item \textbf{RQ1: } What performance does \ourSolution{} achieve in a supervised IO detection task compared to state-of-the-art methods?
    \item \textbf{RQ2: } In a more realistic scenario with limited labeled data, can \ourSolution{} maintain strong predictive performance and outperform the best state-of-the-art method in this data regime?
    \item \textbf{RQ3: } 
    Does a pre-training on a massive dataset enable \ourSolution{} to generalize effectively across unseen IOs?
\end{itemize}

\subsection{Experimental Settings}
\spara{Datasets.}
We conduct our experiments on 6 datasets from \cite{seckin2024labeled}, including IO activities in countries such as UAE, Cuba, Russia, Venezuela, Iran, and China. 
In accordance with previous studies~\cite{luceri2024unmasking}, we select these state-sponsored campaigns for their extensive scale, evident from their size in Table~\ref{tab:dataset_statistics}.
Each country can include multiple campaigns, mirroring real-world situations where both campaign-based and organic conversations from a single country might intersect.

\begin{table}[ht]
\centering\small
\begin{tabular}{lcccc}
\toprule
{\small Country} & {\small Nodes} & {\small Edges} & {\small Homophily} & {\small IO Prop.} \\
\midrule
UAE        & 9242  & 2118684 & 52.8\% & 35.7\% \\
Cuba       & 19822 & 4737374 & 37.1\% & 2.3\%  \\
Russia     & 666   & 10381   & 53.0\% & 38.4\% \\
Venezuela  & 4980  & 56700   & 77.7\% & 10.6\% \\
Iran       & 12977 & 392938  & 81.0\% & 32.2\% \\
China      & 22694 & 410979  & 41.1\% & 3.3\%  \\
\bottomrule
\end{tabular}
\caption{Dataset statistics for different countries. Homophily indicates the percentage of edges connecting nodes of the same class, and IO Prop. indicates the proportion of IO drivers.}
\label{tab:dataset_statistics}
\end{table}

The datasets exhibit different properties in terms of number of nodes, density, edge homophily and label imbalance. We use the class-insensitive edge homophily~\cite{lim2021large} to evaluate the degree of homophily in each graph.
Notably, there is significant variation in homophily across the datasets, ranging from approximately 37\% to 81\%. The presence of heterophilic edges may impact the effectiveness of standard GNN models~\cite{zheng2022graph}.

\begin{table*}[t!]
    \centering\small
    \begin{tabular}{>{\raggedright\arraybackslash}p{1.8cm} >{\raggedright\arraybackslash}p{1.6cm} >{\raggedright\arraybackslash}p{1.45cm} >{\raggedright\arraybackslash}p{1.6cm} >{\raggedright\arraybackslash}p{1.6cm} >{\raggedright\arraybackslash}p{1.6cm} >{\raggedright\arraybackslash}p{1.6cm} >{\raggedright\arraybackslash}p{1.6cm} >{\raggedright\arraybackslash}p{1.0cm}}
        \toprule
        \textbf{Input Features} & \textbf{Model} & \multicolumn{6}{c}{\textbf{Dataset}} \\
        \cmidrule(lr){3-9}
        & & UAE & Cuba & Russia & Venezuela & Iran & China & \textbf{Average} \\
        \midrule
        \multirow{2}{3cm}{\texttt{Graph}} & NodePruning & 84.66{\scriptsize $\pm$ 0.63} & 57.92{\scriptsize $\pm$ 2.07} & 87.65{\scriptsize $\pm$ 1.95} & 95.05{\scriptsize $\pm$ 1.47} & 60.83{\scriptsize $\pm$ 1.09} &  63.66{\scriptsize $\pm$ 0.70} & 74.96 \\
        {} & node2vec+RF &  96.97{\scriptsize $\pm$ 0.42} & \underline{91.53{\scriptsize $\pm$ 1.11}} & 83.43{\scriptsize $\pm$ 3.24} & 90.32{\scriptsize $\pm$ 2.14} & 80.50{\scriptsize $\pm$ 0.60} & \underline{83.89{\scriptsize $\pm$ 1.06}} & 87.77 \\
        \midrule
        \multirow{1}{3cm}{\texttt{Text}} & SBert & 86.23{\scriptsize $\pm$ 0.54} & 90.43{\scriptsize $\pm$ 1.17} & 84.03{\scriptsize $\pm$ 0.91} & 92.74{\scriptsize $\pm$ 1.38} & 85.18{\scriptsize $\pm$ 0.90} & 75.85{\scriptsize $\pm$ 1.53} & 85.74 \\
        \midrule
        \multirow{2}{3cm}{\texttt{Graph}+\texttt{Text}} 
        {} & GNN &  \underline{98.53{\scriptsize $\pm$ 0.13}} & 85.22{\scriptsize $\pm$ 17.9}  & \underline{90.04{\scriptsize $\pm$ 2.66}}  & \underline{97.46{\scriptsize $\pm$ 0.91}}  & \underline{95.12{\scriptsize $\pm$ 0.30}} & 66.42{\scriptsize $\pm$ 21.1} & \underline{88.80} \\
        & \ourSolution{} & \textbf{98.76{\scriptsize $\pm$ 0.14}} & \textbf{98.93{\scriptsize $\pm$ 0.38}}$^*$  & \textbf{92.28{\scriptsize $\pm$ 2.24}}$^*$ & \textbf{99.11{\scriptsize $\pm$ 0.32}}$^*$ & \textbf{96.61{\scriptsize $\pm$ 0.35}}$^*$ & \textbf{92.90{\scriptsize $\pm$ 1.00}}$^*$ & \textbf{96.43} \\
        \bottomrule
    \end{tabular}
    \caption{Results in terms of Macro-F1 for various models across datasets on the task of IO driver detection. The table reports both the average and standard deviation across five different random seeds. The best results are highlighted in bold, while the second-best are underlined. A statistically significant improvement between the best and second-best results (i.e., $p < 0.05$) according to a two-sided t-test is indicated with a star.}
    \label{tab:results}
\end{table*}

\spara{Models.}
To evaluate \ourSolution{} performances, we compare it with state-of-the-art methods for the task of IO detection. 
We select two methods introduced by~\citeauthor{luceri2024unmasking}~(\citeyear{luceri2024unmasking}). 
The first one, \textbf{NodePruning} categorizes a user as an IO driver depending on their eigenvector centrality in the Fused Similarity Network. If this centrality value exceeds a certain threshold, the user is labeled as an IO driver.
The second is based on node2vec (\textbf{Node2vec+RF}), which extracts node embedding based on the local network topology. The node2vec representations of the Fused Similarity Network are then used as input features for a Random Forest classifier to detect whether a user is an IO driver or not. To extend the set of baselines, we also include two more architectures. The first one is a shallow multilayer perceptron (\textbf{MLP}) trained on content-based features (\textbf{SBert+MLP}). Each user is represented by the average of their top 5 most popular tweets' embedding. Those are extracted using a sentence transformer,  SBert~\cite{reimers-2019-sentence-bert}. The last type of architecture we also test is based on GNNs. We test two popular types, GCN and Sage, with either only structural features, only textual features, or a simple concatenation of structural and textual features. 
In this manuscript, we present only the best-performing approach among these, i.e., the one trained with concatenation of text and graph-based features. 


\spara{Implementation details.}
We utilize the PyTorch Geometric~\cite{fey2019fast} library to implement all baselines (except Node Pruning) and \ourSolution{}.
Since IO detection is a binary classification task, we use Macro-F1 as an evaluation metric to address the label imbalance, which can be noted in  Table~\ref{tab:dataset_statistics}.
We follow a 60-20-20 random split strategy to build training, validation, and test sets. We report average and standard deviation across five different seeds to obtain reliable performance estimates.
We optimize each model for 1000 epochs, evaluating the Macro-F1 on the validation set at the end of each epoch, with early stopping implemented after a certain number of epochs of no improvement.
We perform a hyper-parameter search on the learning rate of the Adam optimizer in $\{10^{ -2}, 10^{-3}\}$, on the early stopping number of epochs in $\{20,25,30\}$ and on the number of MLP layers in $\{2,3,4\}$ fed with SBert embeddings. We set the number of latent dimensions to 128 for node2vec, following the configuration of \cite{luceri2024unmasking} and set the same number of hidden neurons for all other methods. 
All models based on GNNs use 2 message-passing layers. 
All neural models use Dropout units with 20\% percentage.
We use a full-batch procedure in Algorithm~\ref{alg:iobuster}, hence $V_{\text{batch}}$ is always equal to $V$.
Experiments are conducted on a DGX Server equipped with 4 NVIDIA Tesla V100 GPU (32GB) and CUDA Version 12.2. 

\begin{figure*}[ht]
    \centering
    \begin{tabular}{ccc}
        \includegraphics[width=0.3\linewidth]{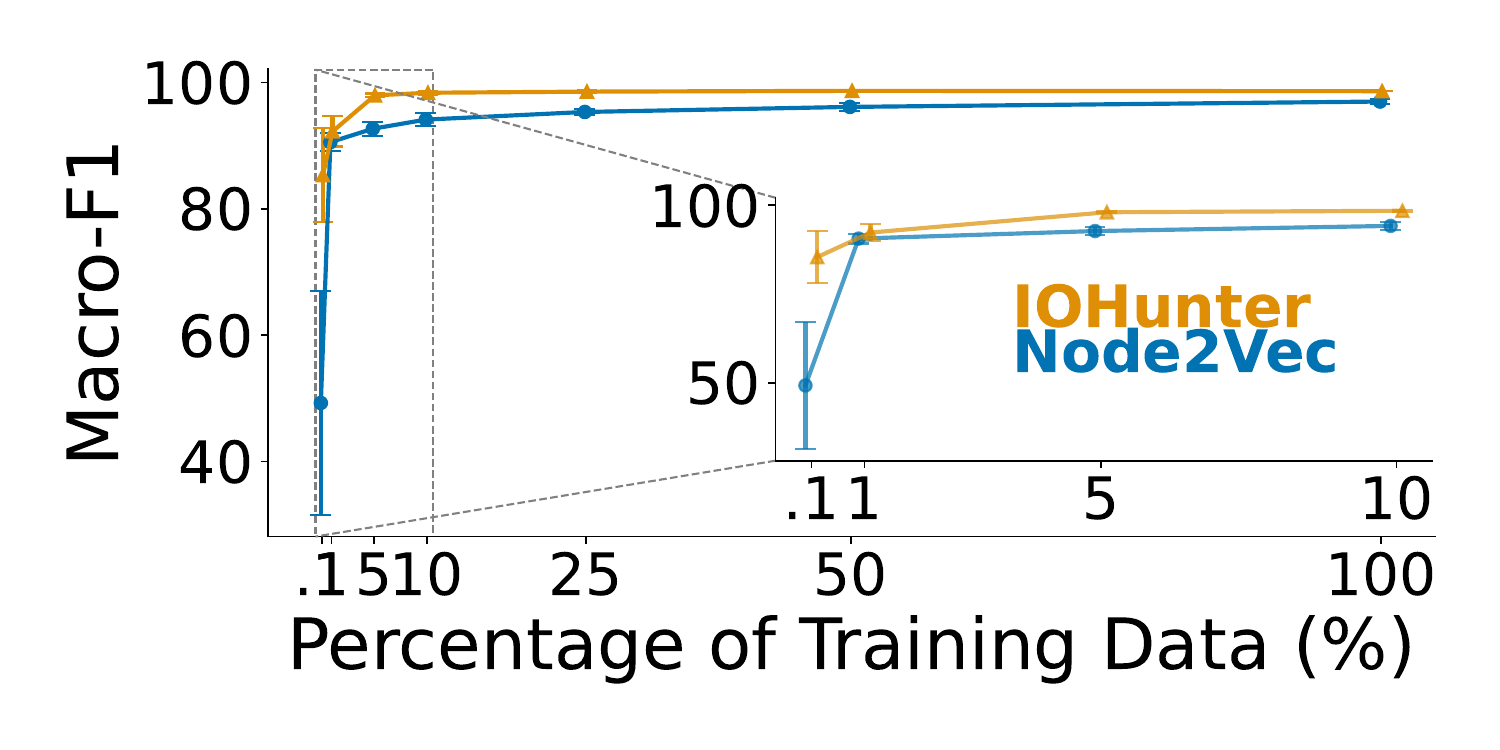} & 
        \includegraphics[width=0.3\linewidth]{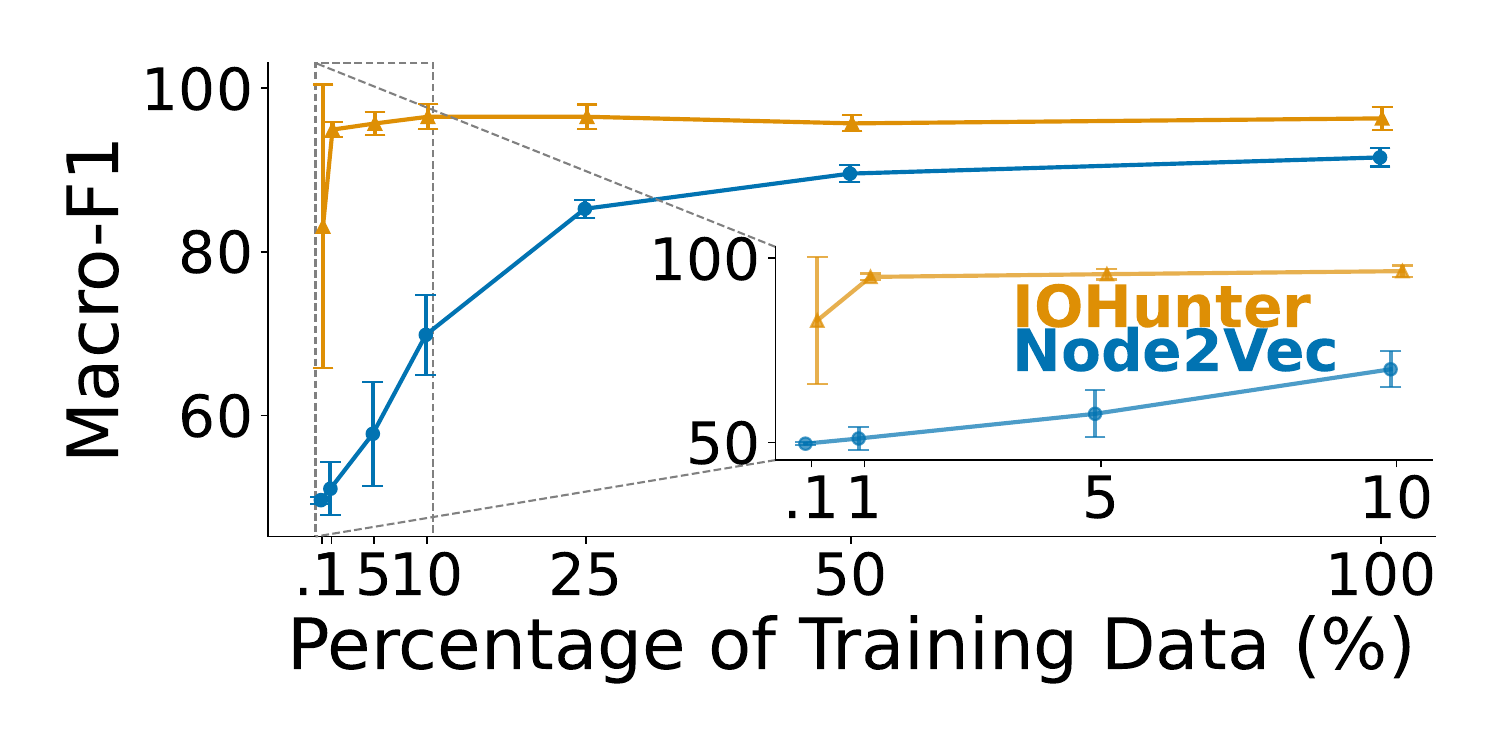} & 
        \includegraphics[width=0.3\linewidth]{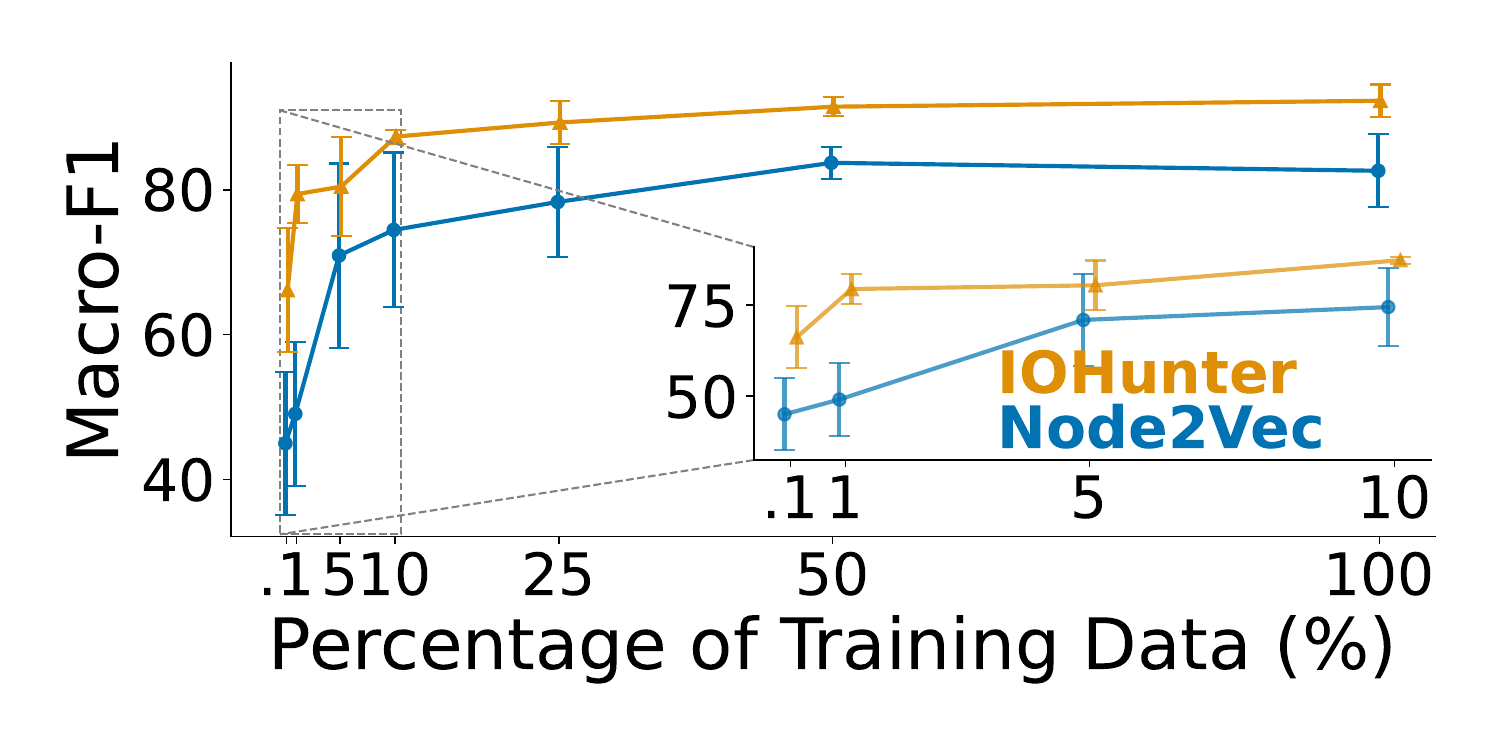} \\
        \textbf{(a)} UAE & \textbf{(b)} Cuba & \textbf{(c)} Russia \\
        \includegraphics[width=0.3\linewidth]{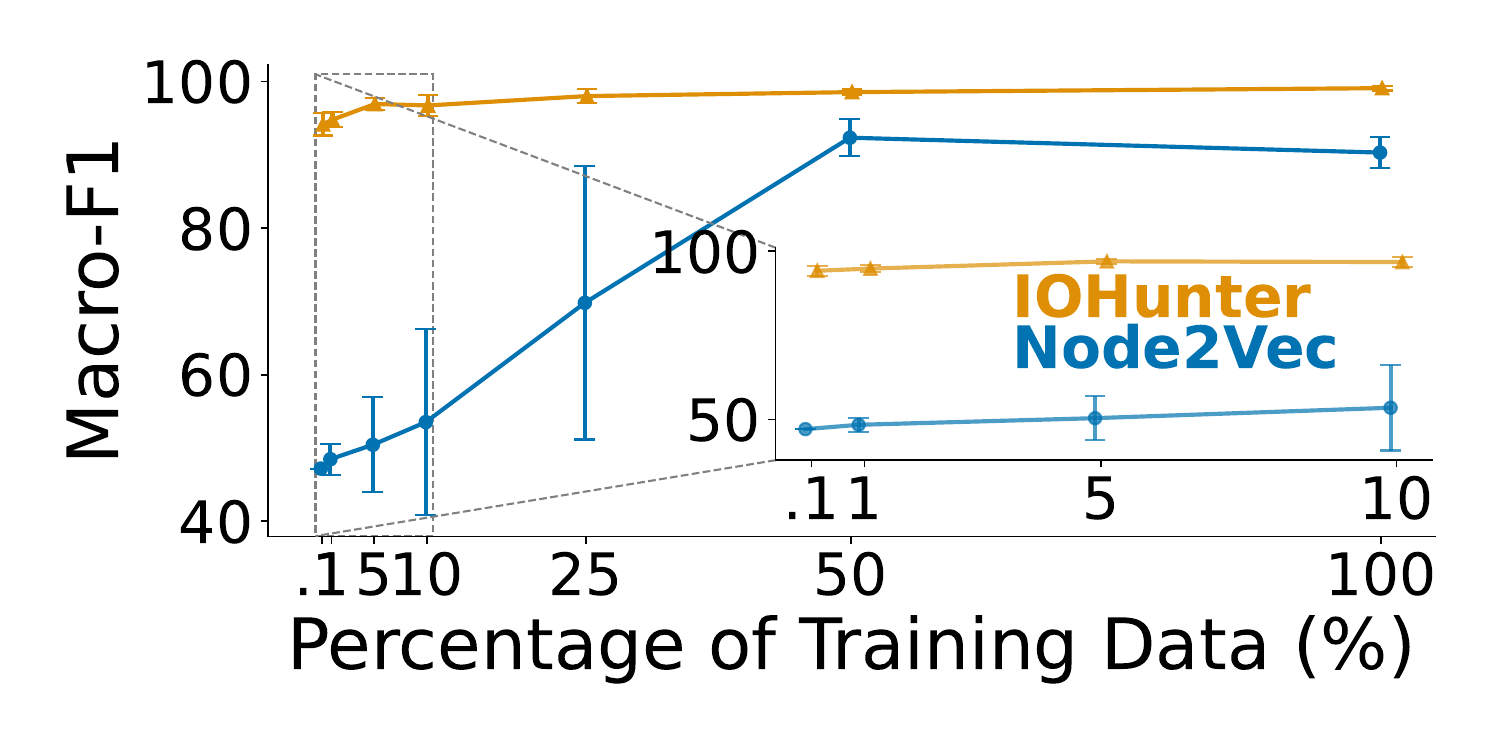} & 
        \includegraphics[width=0.3\linewidth]{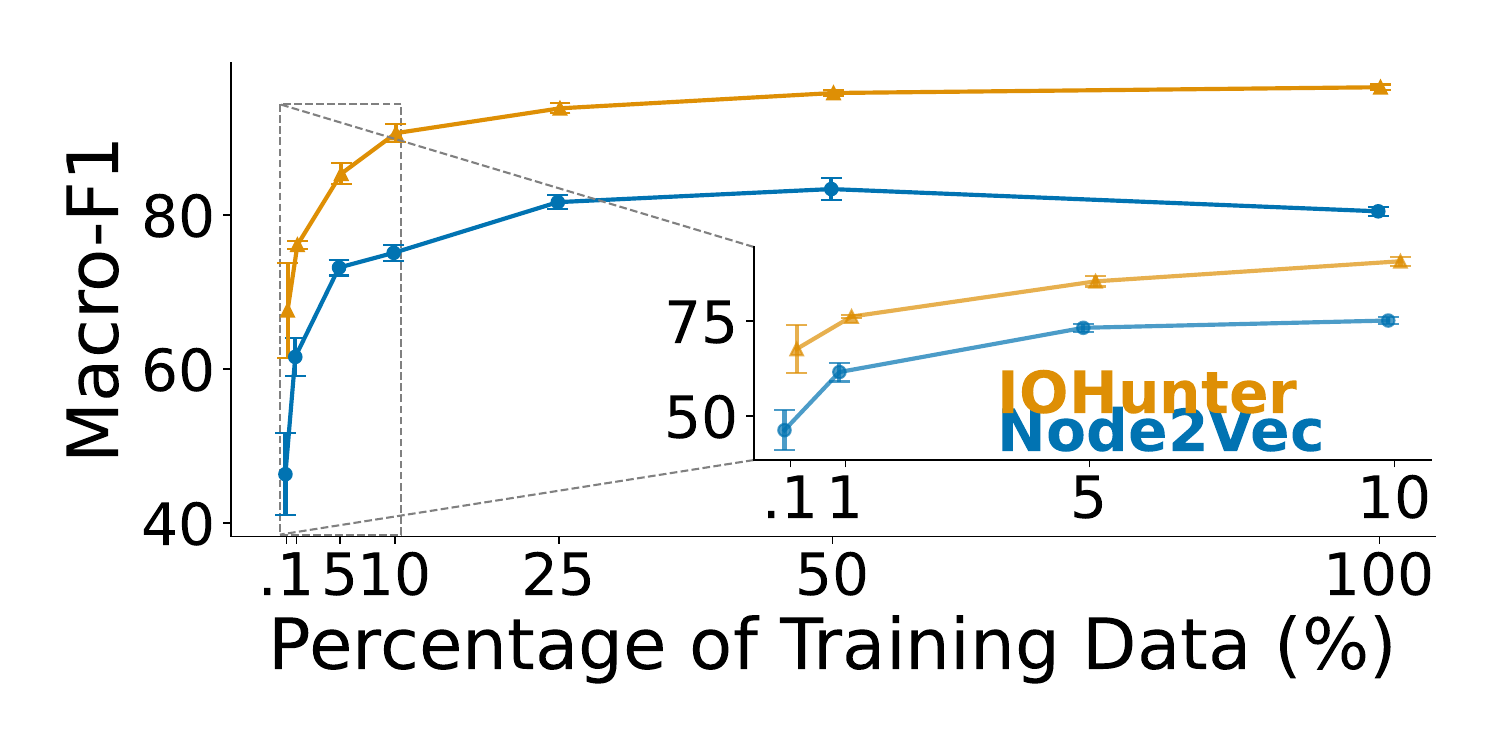} & 
        \includegraphics[width=0.3\linewidth]{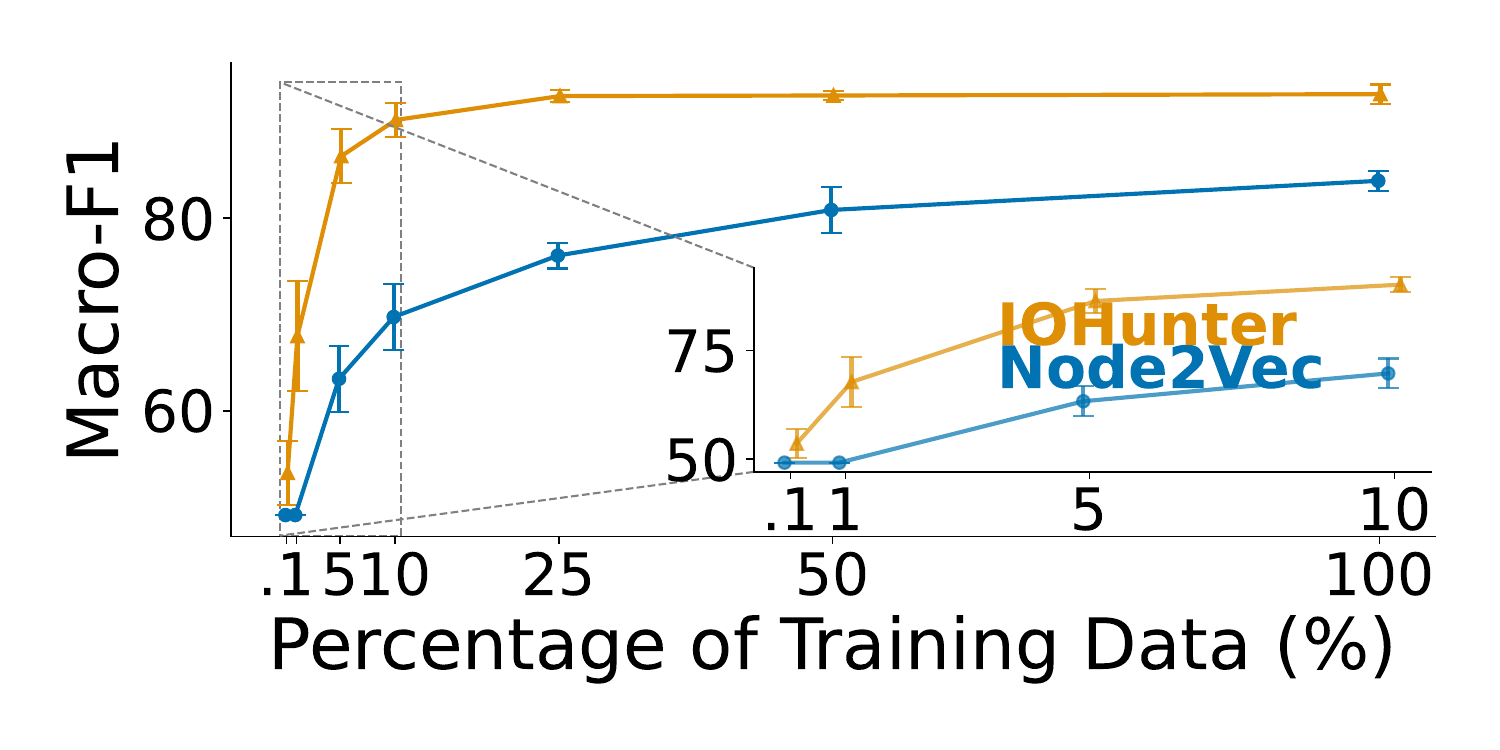} \\
        \textbf{(d)} Venezuela & \textbf{(e)} Iran & \textbf{(f)} China \\
    \end{tabular}
    \caption{Performance of \ourSolution{} and Node2Vec+RF with different amount of training data sparsity.}
    \label{fig:figures_grid}
\end{figure*}

\subsection{Supervised Detection of IOs (RQ1)}

Table \ref{tab:results} shows the performances of \ourSolution{} and compares it with other baselines and existing approaches.
Our proposed method demonstrates a clear and consistent improvement over the state-of-the-art node2vec+RF across all six datasets, with an average percentage gain of 9.25\% in terms of Macro-F1. The performance enhancement is particularly noteworthy, ranging from a gain of +1.8\% on the UAE dataset to a  +20\% on the Iran dataset. This breadth of improvement across diverse datasets underscores the robustness and versatility of our approach, especially in scenarios where traditional methods like node2vec+RF may not fully capture the complexity of the data and diversity of IOs.

A critical factor in \ourSolution{} success is the effective integration of both structural and textual features, which proves essential for achieving competitive results across all datasets. Relying exclusively on either structural or textual features, as evidenced by the inconsistent performance of the NodePruning and SBert models, falls short of delivering reliable outcomes across IOs. 
The need to combine both modalities is paramount, as neither feature set alone can encapsulate the full complexity required for effective user classification. The strength of our approach lies in its ability to seamlessly blend these modalities, which is a key driver behind its superior performance across diverse datasets.

Moreover, the inclusion of a cross-attention mechanism in our model significantly enhances performance compared to a straightforward multi-modal concatenation approach (i.e., GNN in Table~\ref{tab:results}), as evidenced by superior outcomes on 6 out of the 6 datasets. This improvement is not only reflected in higher absolute performance but also in greater stability. The cross-attention mechanism mitigates the variability often observed in models relying on simple concatenation, which can be sensitive to initial conditions such as random seed variations. By effectively capturing the interactions between different data modalities, the cross-attention mechanism enables a more reliable integration, ensuring consistent and robust results.

\subsection{IO Detection under Data Scarcity (RQ2)}
To evaluate the robustness and effectiveness of \ourSolution{}, we conducted a series of experiments designed to test its performance under varying levels of data scarcity. We simulated different degrees of sparsification by downsampling the training set to create sparser versions. Specifically, we train our model using only the 0.1\%, 1\%, 5\%, 10\%, 25\%, and 50\% of the original training data. For each level of sparsification, we trained both \ourSolution{} and node2vec+RF, which is the leading approach in the literature for detecting IOs~\cite{luceri2024unmasking}.
This experimental design allows us to assess the robustness of \ourSolution{} in scenarios where labeled data is extremely limited, a common challenge in real-world applications.

Figure \ref{fig:figures_grid} portrays the results of this analysis.
\ourSolution{} consistently outperforms the node2vec+RF  across all levels of sparsification and across all six datasets. This consistent superiority across various data regimes highlights the robustness of our model and underscores its potential applicability in real-world scenarios where data availability can often be a limiting factor in detecting new, unseen IOs.

In particular, the performance gap between \ourSolution{} and node2vec+RF becomes increasingly pronounced as the training data is reduced, especially in the datasets for Cuba, Russia, Venezuela, and China. For these countries, as the level of sparsification increases, \ourSolution{} maintains a relatively stable performance while the compared model deteriorates significantly. 

Notably, in the UAE, Cuba, and Venezuela datasets, even with as little as 0.1\% of the training data, \ourSolution{} is able to achieve a Macro-F1 score of approximately 80\%. This is a remarkable result, especially given the extremely limited amount of training data. However, it is also worth noting that with such sparse data, there is more significant variability in the results across different random seeds. 

\subsection{Generalizable Cross-IO Detection (RQ3)}
To evaluate the cross-IO generalization capability of \ourSolution{}, we designed an experiment inspired by the principles underlying foundation models in computer vision~\cite{zhai2022scaling, cherti2023reproducible} and natural language processing~\cite{kaplan2020scaling, hernandez2021scaling}. The goal of this experiment is to determine whether pretraining the model on a diverse set of countries can enable it to generalize effectively to a new country for which no training data has been seen during pretraining. Also, we assess whether fine-tuning the pretrained model on a small fraction of data from the test country can further enhance its performance, particularly when labeled data is scarce.

In this experiment, we employ a leave-one-out strategy where the model is pretrained on data from all countries except one, which is reserved as the test country. We repeat this process for each of the six countries: Cuba, Russia, Venezuela, China, United Arab Emirates (UAE), and Iran. Following the pretraining phase, we evaluate the model's performance on the test country both in its pretrained state (referred to as the ``Only PreTrain" strategy) and after fine-tuning it on just 0.1\% of the training data from the test country (referred to as the ``PreTrain \& FineTune on 0.1\%" strategy). The results are then compared against a baseline model that is trained only on 0.1\% of the test country's training data without any pretraining.

\begin{table*}[h!]
    \centering\small
    \begin{tabular}{>{\raggedright\arraybackslash}p{3.85cm} 
                    >{\centering\arraybackslash}p{1.45cm} 
                    >{\centering\arraybackslash}p{1.45cm} 
                    >{\centering\arraybackslash}p{1.45cm} 
                    >{\centering\arraybackslash}p{1.45cm} 
                    >{\centering\arraybackslash}p{1.45cm} 
                    >{\centering\arraybackslash}p{1.45cm} 
                    >{\centering\arraybackslash}p{1.75cm}}
        \toprule
        \multirow{2}{*}{Ours} & \multicolumn{7}{c}{Datasets} \\
        \cmidrule(lr){2-8}
         & UAE & Cuba & Russia & Venezuela & Iran & China & $\Delta$\% No PreTraining \\
        \midrule
        {\small Train on 0.1\%} & 85.33{\scriptsize $\pm$ 7.38} & 83.09{\scriptsize $\pm$ 17.3} & 66.13{\scriptsize $\pm$ 8.52} & \textbf{94.16{\scriptsize $\pm$ 1.51}} & 67.65{\scriptsize $\pm$ 6.20} & 53.55{\scriptsize $\pm$ 3.33} & \_\_\_\_ \\
        \midrule
        \midrule
        {\small Only PreTrain} & 83.93{\scriptsize $\pm$ 5.93} & \underline{89.91{\scriptsize $\pm$ 5.35}} & \underline{79.77{\scriptsize $\pm$ 1.93}} & 90.99{\scriptsize $\pm$ 1.07} & \underline{72.78{\scriptsize $\pm$ 1.43}} & \underline{58.14{\scriptsize $\pm$ 5.89}} & {\scriptsize +}5.69\% \\
        {\small PreTrain \& FineTune on 0.1\%} & \textbf{88.97{\scriptsize $\pm$ 4.37}} & \textbf{91.27{\scriptsize $\pm$ 3.12}} & \textbf{85.09{\scriptsize $\pm$ 2.41}} & \underline{92.10{\scriptsize $\pm$ 1.95}} & \textbf{73.75{\scriptsize $\pm$ 1.27}} &  \textbf{64.88{\scriptsize $\pm$ 1.78}} & {\scriptsize +}10.25\%\\
        \bottomrule
    \end{tabular}
    \caption{Results of \ourSolution{} in terms of Macro-F1 for the cross-country detection task. For each country $c$, \ourSolution{} is pretrained on data from all other countries (i.e., Only PreTraining) and then tested on $c$'s test set. \ourSolution{} can be further fine-tuned on a tiny percentage (0.1\%) of $c$ training set. We compare the model trained on the 0.1\% of $c$'s training set.}
    \label{tab:cross-country-results}
\end{table*}

\begin{table*}[h!]
    \centering\small
    \begin{tabular}{>{\raggedright\arraybackslash}p{3.75cm} 
                    >{\centering\arraybackslash}p{1.45cm} 
                    >{\centering\arraybackslash}p{1.45cm} 
                    >{\centering\arraybackslash}p{1.45cm} 
                    >{\centering\arraybackslash}p{1.45cm} 
                    >{\centering\arraybackslash}p{1.45cm} 
                    >{\centering\arraybackslash}p{1.45cm}
                    >{\centering\arraybackslash}p{1.45cm}}
        \toprule
        \textbf{Model} & UAE & Cuba & Russia & Venezuela & Iran & China & \textbf{Average}\\
        \midrule
        \ourSolution{} w/o Graph     & 98.46 & 96.16 & 89.80 & 86.86 & 91.73 & 57.22 & 86.76\\
        \ourSolution{} w/o Text      & 82.54 & 86.86 & 88.04 & 98.34 & 84.35 & 91.05 & 88.53\\
        \ourSolution{} w/o CrossAttn & 98.53 & 85.22 & 90.04 & 97.46 & 95.12 & 66.42 & 88.80\\
        \ourSolution{}               & \textbf{98.76} & \textbf{98.93} & \textbf{92.28} & \textbf{99.11} & \textbf{96.61} & \textbf{92.90} & \textbf{96.43}\\
        \bottomrule
    \end{tabular}
    \caption{Results of the ablation study on the proposed architecture, i.e. \ourSolution{}. Metrics represent Macro-F1 on the test set. The full configuration of \ourSolution{} consistently achieves the best results, highlighted in bold.}
    \label{tab:ablation-study}
\end{table*}

Cross-country generalization experiments, presented in Table~\ref{tab:cross-country-results}, highlight the significant advantages of our pretraining strategy. The ``Only PreTrain" strategy, which involves pretraining on all countries except the test country, achieves remarkable performance. In four out of the six datasets, this strategy surpasses the performance of a model trained in a fully-supervised fashion on 0.1\% of the test country's training data. This outcome demonstrates the model's strong ability to generalize across different datasets, effectively transferring knowledge from one domain to another. Furthermore, when we apply the ``PreTrain \& FineTune on 0.1\%" strategy, the performance improves even further. Fine-tuning the pretrained model on a tiny percentage of the test country's data leads to an average improvement of more than 10\% in Macro-F1 scores across all countries. This result underscores the power of combining massive pretraining with targeted fine-tuning, as it allows the model to adapt to the characteristics of the new domain with minimal supervision. 

These findings suggest that pretraining on a diverse set of countries not only equips the model with a strong baseline capability for cross-country generalization but also that fine-tuning on a small fraction of data from the target country can yield substantial performance gains. This approach is particularly valuable in scenarios where labeled data is scarce or expensive to obtain. 

\subsection{Ablation Study}
We present the results of experiments designed to evaluate the importance of three key components of \ourSolution{}: the cross-attention mechanism for integrating the two modalities, the graph modality, and the text modality. The findings of this ablation study are summarized in Table~\ref{tab:ablation-study}.

Our analysis shows that using a simple concatenation of the two modalities (denoted as ``\ourSolution{} w/o CrossAttn" in the table) is suboptimal—consistent with the results in Table~\ref{tab:results}—and offers limited advantages compared to a uni-modal classifier based solely on the graph modality (denoted as ``\ourSolution{} w/o Text" in the table). This aligns with prior findings on the challenges of training a multi-modal classifier that consistently outperforms the best uni-modal alternative~\cite{wang2020makes}. Moreover, incorporating the text modality via the cross-attention mechanism (denoted as ``\ourSolution{}" in the table) leads to an approximately 9\% improvement over the graph-only approach. This result highlights the need to use both information sources for an effective IO classifier.

\section{Conclusions}
In this work, we introduced \ourSolution, a novel GFM designed to detect IOs across various social media platforms. By integrating the strengths of LMs and GNNs, our approach leverages both textual and network-based features to accurately identify IO drivers across multiple countries and campaigns. 

\ourSolution{} significantly outperforms state-of-the-art methods in \textit{supervised}, \textit{scarcely-supervised}, and \textit{cross-country} IO settings. This is achieved through the model's ability to generalize across diverse and evolving contexts, enabled by pretraining on extensive, multi-country datasets and fine-tuning with minimal labeled data. 
The robustness of \ourSolution{} in scenarios with limited data availability underscores its practical applicability in real-world settings, where access to labeled data is often limited. Additionally, the cross-country generalization experiments highlight the model's potential to transfer knowledge between different geopolitical contexts.

Moving forward, our work paves the way for the development of more sophisticated GFMs tailored to more graph-related tasks. Future research could explore the application of \ourSolution{} to domains where detecting coordinated, malicious activities is critical, as well as the integration of additional data modalities to further improve detection capabilities. 
Ultimately, \ourSolution{} represents a significant step toward safeguarding the integrity of online discourse by providing a scalable, adaptable, and highly effective solution for uncovering information operations on social media platforms.

\section*{Acknowledgements}
Work supported in part by DARPA (contract \#HR001121C0169). MM acknowledges partial support by the SERICS project (PE00000014) under the NRRP MUR program funded by the EU - NGEU.

\bibliography{references}

\end{document}